\documentclass[letterpaper, 10 pt, conference]{ieeeconf}  

\IEEEoverridecommandlockouts                              
\usepackage{xcolor}
\usepackage{subfigure}
\usepackage{graphicx}
\usepackage[frozencache]{minted}
\setminted{fontsize=\footnotesize,baselinestretch=.97,linenos,frame=lines,xleftmargin=6pt,numbersep=3pt,mathescape=true,escapeinside=^^,bgcolor=bg}
\definecolor{bg}{rgb}{0.97,0.97,0.97}
\newminted{chapel}{}

\title{\LARGE \bf
Paving the way for Distributed Non-Blocking Algorithms and \\ Data Structures in the Partitioned Global Address Space model}

\author{Garvit Dewan, Indian Institute of Technology Roorkee, gdewan@cs.iitr.ac.in \\ Louis Jenkins, University of Rochester, louis.jenkins@rochester.edu\thanks{This work was supported by the US Department of Energy Computational Science Graduate Fellowship (grant DE-SC0020347). Special thanks to Cray, a Hewlett-Packard Enterprise Company, for providing access to the compute resources utilized during development.}}

\begin{document}

\maketitle
\thispagestyle{empty}
\pagestyle{empty}

\begin{abstract}

The partitioned global address space memory model has bridged the gap between
shared and distributed memory, and with this bridge comes the
ability to adapt shared memory concepts, such as non-blocking
programming, to distributed systems such as supercomputers. 
To enable non-blocking algorithms, we present ways to perform
scalable atomic operations on objects in remote memory
via remote direct memory access and pointer
compression. As a solution to the problem of concurrent-safe
reclamation of memory in a distributed system, we adapt
Epoch-Based Memory Reclamation to distributed memory and
implement it such that it supports global-view programming.
This construct is designed and implemented for the Chapel
programming language but can be adapted and generalized to
work on other languages and libraries.

\end{abstract}

\section{INTRODUCTION}


In synchronized data structures and algorithms, there are many pitfalls that programmers can fall into, such as deadlock, livelock, and priority inversion~\cite{mAlgorithmsScalableSynchronization1991}.
Non-blocking data structures and algorithms provide many benefits over their synchronized counterparts, such as providing guarantees on liveness, which is a property on how threads make progress throughout a system. One such liveness property is \textit{obstruction-freedom}~\cite{herlihyObstructionfreeSynchronizationDoubleended2003}, which states that threads are guaranteed to complete their operation so long as they are not obstructed by some other thread, such as by executing in isolation; in \textit{lock-freedom}~\cite{massalin1992lock}, at least one thread is guaranteed to progress and succeed in a bounded number of steps; in \textit{wait-freedom}~\cite{10.1145/114005.102808}, all threads are guaranteed to progress and succeed in a bounded number of steps. These properties are attainable in shared-memory, with decades of research available on non-blocking data structures for shared-memory. However, to the authors' knowledge, there are not many, if at all, for distributed memory. 

The \textit{Partitioned Global Address Space (PGAS)} memory model offers an abstraction of distributed memory systems in a way that allows them to have very similar semantics to that of shared-memory. For example, PUTs and GETs, which are \textit{remote-direct memory access (RDMA)} operations that remotely write and read values in memory without the intervention of the CPU, are analogous to shared-memory load and store operations, and can even be modeled as such ~\cite{hayashiLLVMbasedCommunicationOptimizations2015}. RDMA atomic operations, which are entirely handled by the network interface controller (NIC) in high-performance computing networks such as InfiniBand and Gemini/Aries, allow for extremely low latency atomic operations that are in the ballpark of mere microseconds. The exploration of the application of non-blocking algorithms and data structures to PGAS is enticing, but there are roadblocks and hurdles that must be dealt with first. For example, the Chapel programming language lacks native support for atomics on arbitrary objects such as class instances, which is necessary for any non-blocking algorithm. As well, the issue of concurrent-safe memory-reclamation, that is, the reclamation of memory when arbitrarily many threads could be accessing said memory at any given time, is a real problem in shared-memory where multiple solutions exist~\cite{lockfreerefcounting, michaelHazardPointersSafe2004,wenIntervalbasedMemoryReclamation2018, hartPerformanceMemoryReclamation2007a}.


This work provides \texttt{EpochManager} and \texttt{LocalEpochManager}, which are based on Epoch-Based Reclamation (EBR) ~\cite{fraser2004practical}. Both serve as pseudo garbage collection mechanisms that scale not only in shared-memory but in distributed memory as well. Also provided is \texttt{AtomicObject}, and the local-optimized variant \texttt{LocalAtomicObject}. Both provide atomic operations on class instances and provide optional ABA-protection. The former is designed to support RDMA atomics on class instances, making it possible for some truly scalable algorithms and data structures.

\section{Design \& Implementation}

In the development of the \texttt{EpochManager}, there were prerequisites that needed to be addressed, such as the need for atomic operations on class instances. Only after overcoming these hurdles is it possible to create the infrastructure and building blocks necessary for creating non-blocking algorithms in both shared and distributed memory. 

\subsection{Atomic Objects}

In Chapel, \textit{atomic} operations, which are operations that appear to take place all at once from any other task's point of view, are defined only on \textit{bool}, \textit{int}, \textit{uint} and \textit{real} primitive types. As of today, there is no official support for atomic operations on class instances, which are represented as \textit{widened} pointers that contain not only the 64-bit virtual address but 64 bits of locality information, comprising a 128-bit structure. Chapel has not implemented support for atomics on class instances due to portability challenges, creating significant obstacles to create even the most primitive of non-blocking data structures, such as queues, stacks, and linked lists. Furthermore, in Chapel, atomic operations over the network rely upon \textit{Remote Direct Memory Access (RDMA)}, which currently only supports atomic operations up to 64-bit.

In the initial prototype, which has been adapted into its independent module, called the \texttt{LocalAtomicObject}, the locality information is ignored, and it maintains an atomic holding only the 64-bit virtual address. As \texttt{LocalAtomicObject} will only work in a shared-memory context, the \texttt{GlobalAtomicObject} offers \textit{pointer compression}, which is designed to take advantage of the fact that currently, processors only use the lowest 48-bits for the virtual address, enabling the encoding of 16-bits of locality information in the 64-bit pointers. This approach will only work in distributed setups with fewer than $2^{16}$ compute nodes, which consequently enables RDMA atomics on Cray Aries.\footnote{RDMA atomics are not yet possible on InfiniBand networks due to a lack of current support in Chapel's implementation.} In the event that more than $2^{16}$ compute nodes are used, the implementation will fall back to using x86 \textit{CMPXCHG16B} instruction\footnote{The equivalent load-linked/store-conditional instructions can be used on ARM.}, also known as the `Double-word-Compare-and-Swap' (DCAS) operation, which can atomically update both the virtual address as well as the 64-bits of locality information. Unfortunately, this demotes atomic operations on remote memory from RDMA atomics, which take around a microsecond to complete and do not require the intervention of the CPU, to using active messages, which are entirely handled by the progress thread of the recipient compute node.
As \texttt{shared} type is already wrapped in a record and is larger than 64-bits, \texttt{owned} type is statically managed and cannot be tracked without significant rework to the type, and \texttt{borrowed} types are explicitly tracked by the compiler making it difficult to track without some significant rework, support is currently restricted to \texttt{unmanaged} class instances. Support for \texttt{owned} and \texttt{borrowed} types is planned as a future work.

Another problem that had to be overcome was the \textit{ABA} problem. The ABA problem occurs in scenarios where one has at least two threads, and typically arises when performing a compare-and-swap operation. In one such scenario, consider an atomic linked list where one has multiple threads, where a thread $\tau_1$ reads from the head of the list and receives the node with virtual address $\alpha$. Imagine that $\tau_1$ gets preempted, and some thread $\tau_2$ also reads the head of the list, atomically moves the head of the list forward, and deletes the node such that $\alpha$ is put back on some free-list. Later, some other thread $\tau_3$ allocates a new node which happens to have the same address $\alpha$ and atomically inserts this at the head of the list. Now, $\tau_1$ wakes up and incorrectly succeeds in its atomic exchange, despite the fact that the head of the list has changed. There are two known ways to solve the ABA problem, and they are to either use a concurrent memory reclamation system, in which is currently being built and leads to a chicken-and-egg paradox, and using a DCAS, where a 64-bit counter is held adjacent to the 64-bit word being atomically updated. In the DCAS approach, the counter gets incremented after each ABA-dependent operation, which causes a DCAS to fail even in the event of the ABA problem, since the 64-bit counter will have changed. The \texttt{AtomicObject} and \texttt{LocalAtomicObject} provide a 128-bit wrapper for 64-bit types, called \texttt{ABA} where such a 64-bit counter is held adjacent to the 64-bit virtual address, which in conjunction with pointer compression can provide both ABA-free atomic operations on remote objects, albeit using remote execution rather than RDMA. Each operation has an ABA variant, which includes the suffix `ABA', that will take into account the 64-bit counter, but the advanced user is free to use both ABA and normal variants interchangeably. Due to Chapel's \texttt{forwarding} decorator, it is possible to use the \texttt{ABA} in a seamless manner as if it were the type it is wrapping, as all methods and field accesses will forward to the underlying instance. Example usage of \texttt{AtomicObject}, implementing a \texttt{push} operation of the Trieber Stack~\cite{hendlerScalableLockfreeStack2004}, can be seen in Listing~\ref{lst:atomicobject}.


\begin{listing}
\begin{minted}{chapel}
proc LockFreeStack.push(newObj : T) {
    var node = new unmanaged Node(newObj);
    do {
      var oldHead = head.readABA();
      node.next = oldHead.getObject();
    } while(!head.compareAndSwapABA(oldHead, node));
}
\end{minted}
\caption{Example usage of AtomicObject}
\label{lst:atomicobject}
\end{listing}

\subsection{Epoch Based Reclamation (EBR)}
    
\begin{figure*}
    \centering
    \includegraphics[scale=0.75]{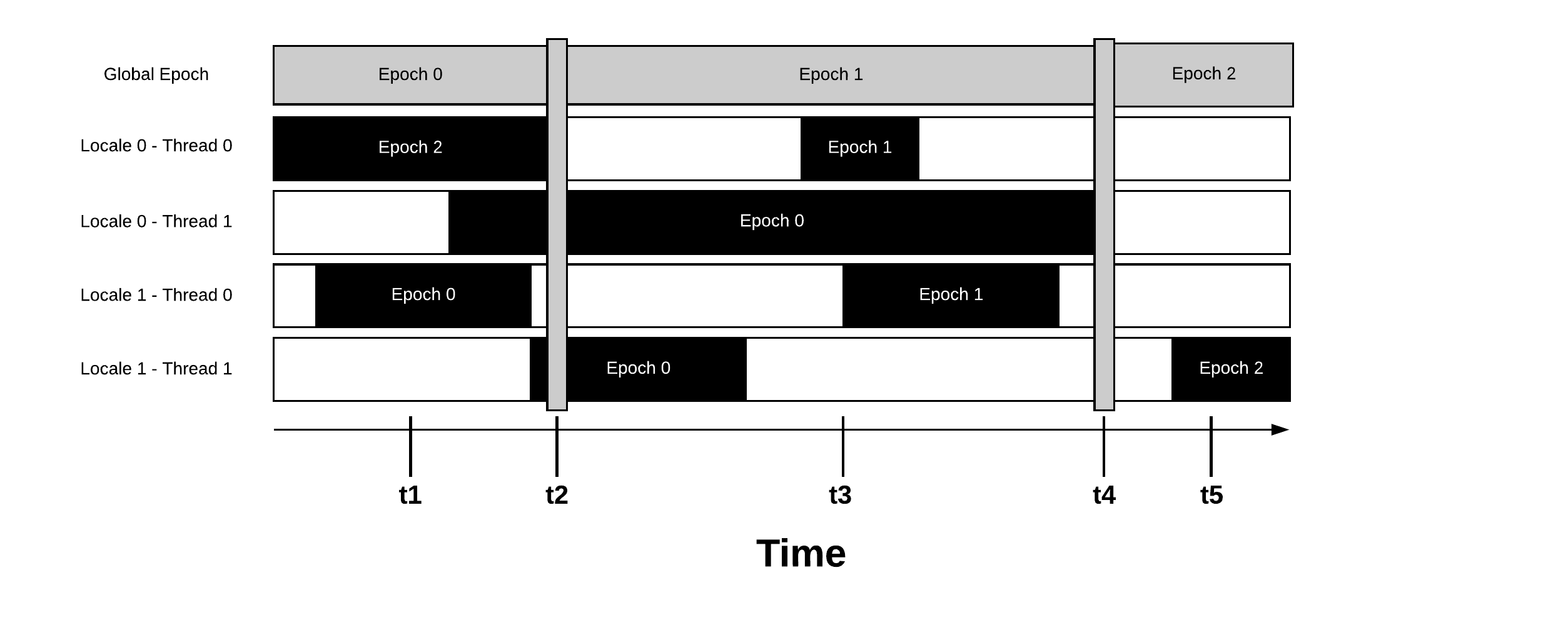}
    \caption{Illustration of Epoch-Based Reclamation, adapted to distributed memory. Given two locales with two threads each, the illustration begins in epoch 0. At time $t_1$, the global epoch is 0 but Locale 0 -- Thread 0 is still in epoch 2, preventing an advancement to epoch 1. At time $t_2$, the global epoch is safely advanced to epoch 1 after Locale 0 -- Thread 0 becomes quiescent; even though Locale 1 -- Thread 0 and Locale 0 -- Thread 1 are in Epoch 0, it is still safe to advance the epoch as epoch 0 is only one epoch behind. Locale 0 -- Thread 1 remains in epoch 0 from $t_1$ through $t_3$, preventing the epoch from advancing until $t_4$, where the global epoch is then advanced to epoch 2. At $t_5$ and onward, the algorithm continues much the same. }
    \label{fig:ebr}
\end{figure*}

It was essential to make the \texttt{EpochManager} non-blocking so to not weaken the non-blocking guarantees of the data structures that employ it, or at least not too much. The three non-blocking guarantees from weakest to strongest are as follows: Obstruction-Freedom, which ensures that if a thread runs in isolation, that is the thread does not have its progress obstructed by any other thread, it will complete in a bounded number of steps; Lock-Freedom, which ensures that at least one thread must complete within a bounded number of steps even when obstructed; Wait-Freedom, which ensures that \textit{all} threads must complete within a bounded number of steps, regardless of obstruction. The \texttt{EpochManager} has been made \textit{lock-free}. 

Epoch-Based Reclamation (EBR) is a concurrent-safe memory reclamation system that utilizes \textit{epochs}, which are descriptors for a specific period of time, to determine the quiescence of objects and determine when they are safe to be reclaimed. Concurrent-safe memory reclamation is a non-trivial problem and is at the very root of non-blocking algorithms and data structures. The problem presented by concurrent access is that it is not easy to know whether or not a thread is accessing data we are interested in deleting, and naively deallocating data can result in undefined behavior from a use-after-free error. That is, once an object is freed, it is normally placed on some type of free-list where it can be used in some future allocation, which can cause data corruption in the case of arbitrary writes or segmentation faults in the case of dereferencing pointers. Epoch-Based Reclamation combats this issue by utilizing epochs. EBR tracks the epochs that participating threads are in, where each participating thread must enter an epoch before accessing data, and must leave the epoch afterwards. Generally, if a thread is not in an epoch, it is considered quiescent in that it no longer has access to the objects we are interested in at that given moment. A thread inside of an epoch may or may not be accessing the object at that given time, but out of safety, the deallocation of said objects is deferred until later. 

To delete an object, it is first \textit{logically} removed from the data structure from which it is accessible. An example of logical removal would be the removal of a node from a linked list. The logically removed object is then put in a \textit{limbo list}, which is a list of objects to be reclaimed, associated with a given epoch. More formally, an object $o$ that is associated with an epoch $e$ must not be deleted until it is certain that no thread is in epoch $e$. The only \textit{hazard} in concurrent memory reclamation occurs when another thread is accessing it while it is being deleted, but the logical removal of $o$ entirely removes it from the data structure, and so only threads that have had access prior to the removal can access $o$. Eventually, once the epoch has been advanced to $e + 1$, which occurs after all threads are guaranteed to be either inactive or in at least epoch $e$ and \textit{not} epoch $e - 1$, it is safe to delete the objects in the limbo list for $e - 1$. Note that $o$ is not reclaimed at this point. Instead, the epoch must advance once more, after which there is utmost certainty that $o$ can safely be reclaimed as all participating threads were quiescent after the logical removal of $o$, and since $o$ is no longer accessible from the current epoch. An illustration of Epoch-Based Reclamation can be seen in Figure~\ref{fig:ebr}.

\subsection{Epoch Manager}

\begin{figure*}
    \centering
    \includegraphics[scale=0.66]{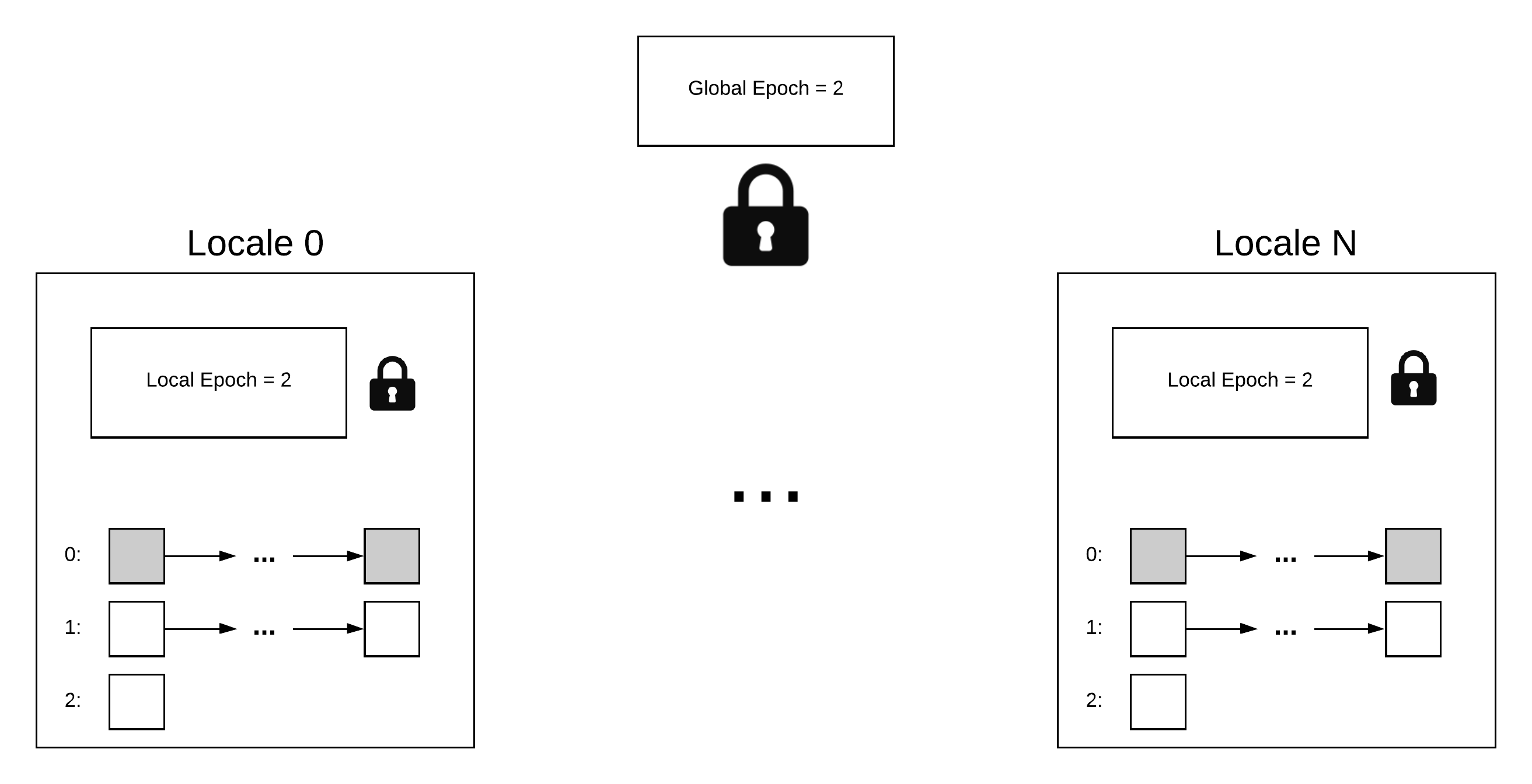}
    \caption{Illustration of EpochManager when global and local epoch is 2. Each locale manages its own privatized instance, where all accesses are directed to. Limbo list 0, shaded gray, is free to be reclaimed as there is a guarantee that no active task will be in epoch $e - 2$; limbo list 1 is not yet reclaimed, and limbo list 2 becomes the current that all new reclaimed objects will be added to. Local Epoch acts as a locale-private cache for the current epoch, reducing necessary communication. An atomic flag must also be acquired to advance the epoch, first locally and then globally, where in failing to do so will cause the attempting thread to back out (non-blocking).}
    \label{fig:epochmanager}
\end{figure*}

The \texttt{EpochManager} is built on top of the notion of epoch-based reclamation and limbo lists, in that objects that are marked for deletion during an epoch are held in limbo until they are safe to be deleted. To implement the limbo lists, it was necessary to implement a non-blocking data structure that was optimized not only for concurrent insertion, but bulk removal, as all objects in the limbo list are deleted at once, and not incrementally. The limbo list can be viewed as having two phases--- an insertion phase, which is entirely concurrent, and a deletion phase, which both occur at disjoint times. A somewhat novel but simple data structure has been designed to significantly reduce overall latency to the point that deferring an object for deletion has been made entirely wait-free during the insertion phase and during the deletion phase, and both are handled in one atomic exchange, shown in Listing~\ref{lst:limbo_list}. Nodes are recycled using a lock-free stack~\cite{hendlerScalableLockfreeStack2004} and the ABA-protection provided by the \texttt{AtomicObject}.

\begin{listing}
\begin{minted}{chapel}
proc push(obj : unmanaged object?) {
    var node = recycleNode(obj);
    var oldHead = _head.exchange(node);
    node.next = oldHead;
}
proc pop() {
    return _head.exchange(nil);
}
\end{minted}
\caption{Wait-Free Limbo List.}
\label{lst:limbo_list}
\end{listing}

The \texttt{EpochManager} is \textit{privatized}, in that an instance of the \texttt{EpochManager} is created and maintained on each locale, and all accesses the \texttt{EpochManager} are \textit{forwarded}, such as the case for field accesses or method invocations, to the instance that is local to that locale. That is, even though the \texttt{EpochManager} can be used in distributed contexts, such as in distributed parallel \texttt{forall} loops, or inside of remote-procedure call (RPC) \texttt{on} statements, all accesses are guaranteed to respect locality. This is achieved by \textit{remote-value forwarding} and \textit{record-wrapping}, where a record which holds data required to lookup the instance itself is based by-value, and not by-reference as is the default in Chapel when it comes to \texttt{forall} statements. This allows for zero-communication when acquiring the privatized instance. This results in a massive speedup, since replication across locales cuts down all unnecessary communication and allows the caching of data or even keeping locale-specific instances of data, and the record-wrapping eliminates an additional round-trip communication required to obtain the metadata needed to find the privatized instance. In practice, this has been observed by the authors to allow distributed objects to no longer be communication bound, that is bound the available bandwidth and latency of the network, and allows for some truly scalable algorithms. This technology is not new, either, as it has been used in previous works to create distributed data structures~\cite{jenkinsChapelAggregationLibrary2018,jenkinsChapelGraphLibrary2019,jenkinsChapelHyperGraphLibrary2018,jenkinsRCUArrayRCULikeParallelSafe2018}, and is used as the backbone for Chapel's arrays, domains, and distributions. An illustration is provided in Figure~\ref{fig:epochmanager}.

An array of three limbo lists are maintained on each locale in the privatized instance, representing the possible epochs that any given thread can be in, which are $e-1$, $e$, and $e+1$. Each locale caches its epoch, which is used when deciding which limbo list to defer deletion of objects to. When it is time to update the global epoch, a task gets elected. In this case, the election is handled in a first-come-first-serve nature via a local atomic flag \texttt{is\_setting\_epoch} for their locale, and then for the locales that the \texttt{EpochManager} is distributed over. This has the effect of stemming off unnecessary amounts of communication that would arise if multiple tasks across multiple locales attempted to update the global epoch at the same time, as only one of them can succeed in doing so. As each locale has its own individual instance, a class instance wraps the global epoch itself so that there is a single centralized and coherent epoch that all locales can come to a consensus on. 

The \texttt{EpochManager} creates a set of \textit{tokens}, which are class instances that keep track of the epoch that a task is currently engaged in. Before a task is free to access a data structure that is protected by the epoch-based reclamation provided by the \texttt{EpochManager}, it must first \textit{register} and obtain one of these tokens. When they are finished, they must \textit{unregister} and relinquish them. Two separate lists are maintained for tokens--- one which keeps track of free tokens, used when registering and unregistering, and one which is a list of all allocated tokens, which is used to scan the minimum epoch. Once registered, the token is not yet in an epoch, and in fact can be used to perform multiple operations in the same task as an optimization. A token must be \textit{pinned} and \textit{unpinned} just like it must be registered and unregistered, where pinning enters the current epoch, and unpinning exits the current epoch. When an object is to be deleted, it is always added to the current epoch associated with the token. The token is itself wrapped in a managed class so that when it goes out of scope, the token can automatically be unregistered. This is particularly useful while using task-private variable intents on \texttt{forall} loops, as shown below.

\begin{listing}[H]
\begin{minted}{chapel}
var em = new EpochManager();
// Serial and Shared Memory
var tok = em.register();
tok.pin();
tok.unpin();
tok.unregister();

// Parallel and Distributed (forall)...
forall x in X with (var tok = em.register()) {
   tok.pin();
   tok.deferDelete(x);
   tok.unpin();
} // automatic unregister
em.clear(); // Reclaim everything at once. 
\end{minted}
\caption{Example usage of \texttt{EpochManager}.}
\end{listing}

The \texttt{EpochManager} will not advance the epoch on its own, and requires user intervention to do so. The user is free to \texttt{tryReclaim}, which attempts to advance the epoch if and only if no token on any other locale is in a previous epoch. As well, since the objects to be deleted can be remote, and since remote deallocation would result in RPC, a scatter list is constructed that sorts objects by the locales they are allocated on, significantly cutting down unnecessary communication. The \texttt{tryReclaim} method is intended to be invoked on the token or \texttt{EpochManager}. It is a global operation, optimized such that if another task is attempting to update the epoch on the current locale, other tasks will swiftly return, without much wasted effort; if another task is attempting to update the global epoch, it will also return after clearing the local flag. The \texttt{clear} method is intended to be invoked directly on the \texttt{EpochManager} and performs the same action as \texttt{tryReclaim} with the exception that it will always reclaim all objects across all epochs, and should be called when there is a guarantee that no other thread is interacting with the \texttt{EpochManager}.

\begin{listing}
\begin{minted}[fontsize=\scriptsize]{chapel}
proc tryReclaim() {
  if (is_setting_epoch.testAndSet()) then return;
  if (global_epoch.is_setting_epoch.testAndSet()) {
    is_setting_epoch.clear();
    return;
  }
    
  // Is it safe to reclaim across all locales?
  const this_epoch = global_epoch.read();
  var safeToReclaim = true;
  coforall loc in Locales with (&& reduce safeToReclaim) 
    do on loc {
    var _this = getPrivatizedInstance();
    for tok in _this.allocated_list {
      var local_epoch = tok.local_epoch.read();
      if (local_epoch != 0 && local_epoch != this_epoch) {
        safeToReclaim = false;
        break;
      }
    }
  }
  
  if safeToReclaim {
    const new_epoch = (current_global_epoch % 3) + 1;
    global_epoch.write(new_epoch);
    coforall loc in Locales do on loc {
      // Update each locale's epoch
      var _this = getPrivatizedInstance();
      _this.locale_epoch.write(new_epoch);

      const reclaim_epoch = _this.getReclaimEpoch();
      var reclaim_limbo_list =
          _this.limbo_list[reclaim_epoch];
      var head = reclaim_limbo_list.pop();

      while (head != nil) {
        var obj = head.val;
        var next = head.next;
        // Scatter objects to their locale
        _this.objsToDelete[obj.locale.id].append(obj);
        delete head;
        head = next;
      }
      coforall loc in Locales do on loc {
        // Bulk transfer and delete
        var ourObjs =
            _this.objsToDelete[here.id].getArray();
        delete ourObjs;
      }
      
      // Clear scatter list
      forall i in LocaleSpace do
        _this.objsToDelete[i].clear();
    }
  }
  global_epoch.is_setting_epoch.clear();
  is_setting_epoch.clear();
}
\end{minted}
\caption{Implementation of \texttt{tryReclaim}.}
\end{listing}

The \texttt{LocalEpochManager} is a shared-memory optimized variant that functions in a similar way to the \texttt{EpochManager}, but differs in that it lacks global epoch and does not take remote objects into consideration when being used, speeding up computations that do not require epoch-based reclamation support across multiple locales.

\section{Performance Evaluation}

\begin{figure}[!tbp]
  \centering
  \begin{minipage}[b]{0.45\textwidth}
    \includegraphics[width=\textwidth]{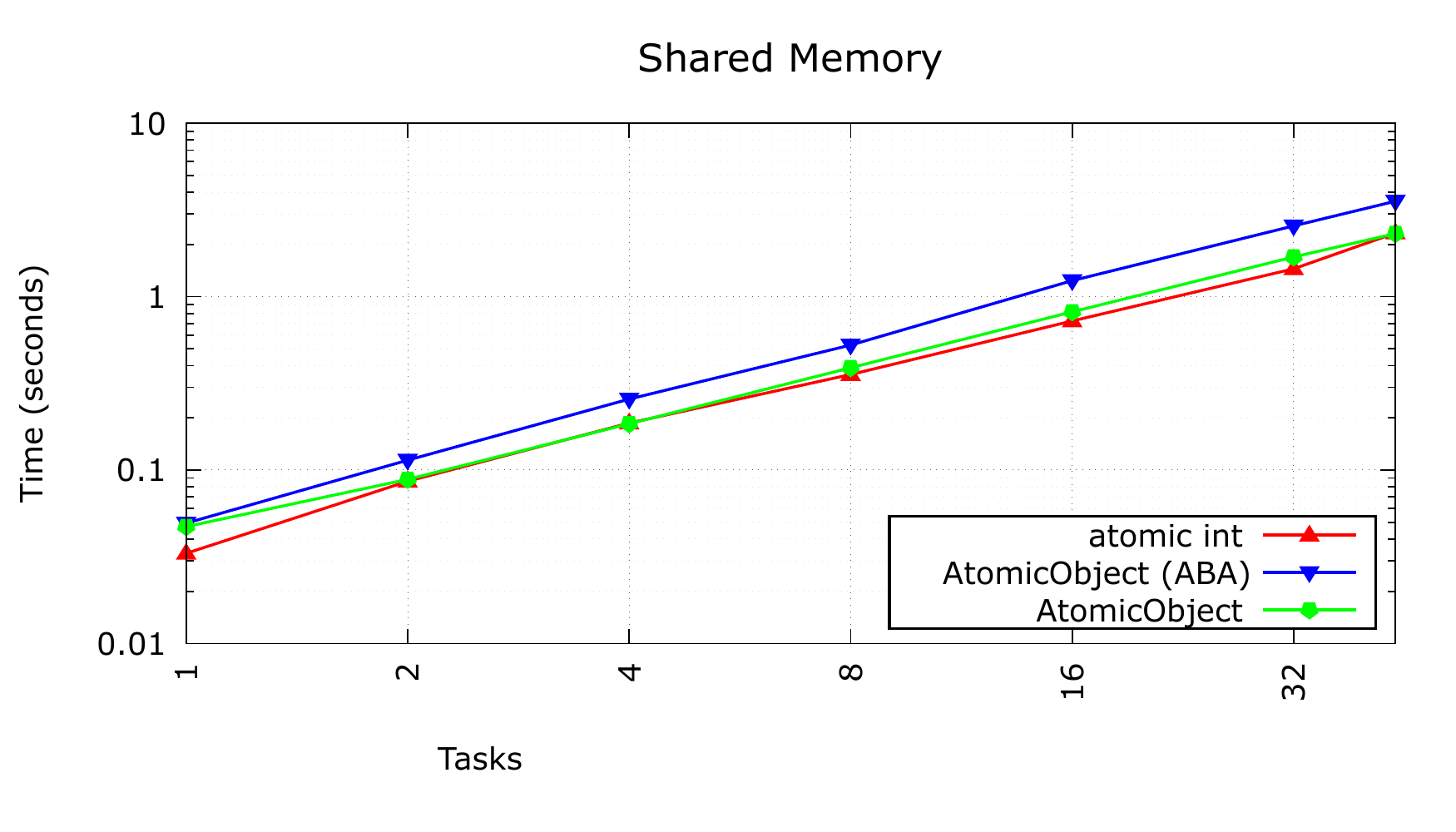}
  \end{minipage}
  \hfill
  \begin{minipage}[b]{0.45\textwidth}
    \includegraphics[width=\textwidth]{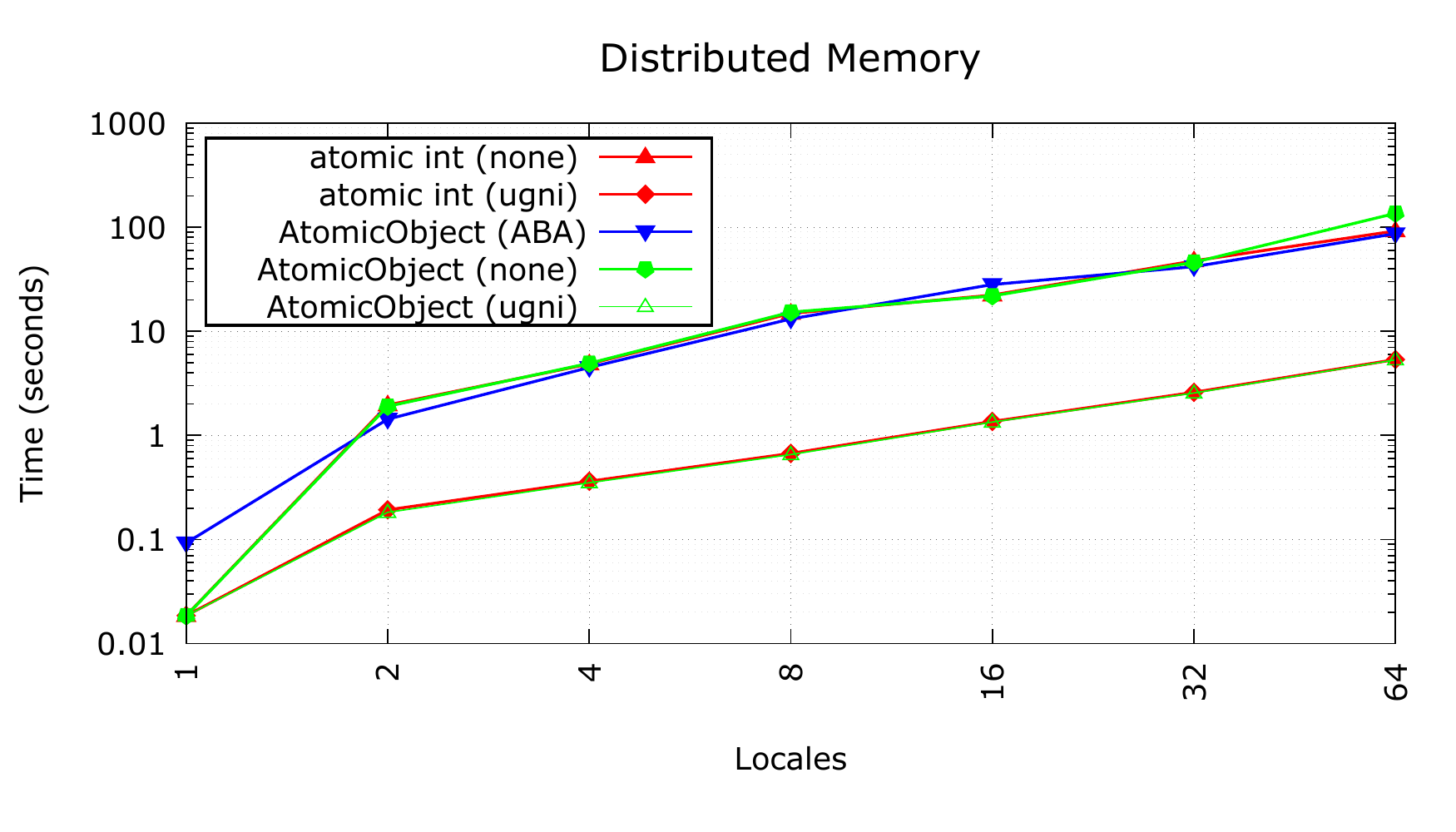}
  \end{minipage}
  \caption{\texttt{AtomicObject} vs \texttt{atomic int}}
  \label{fig:atomics}
\end{figure}

\begin{figure}
    \centering
    \includegraphics[width=0.45\textwidth]{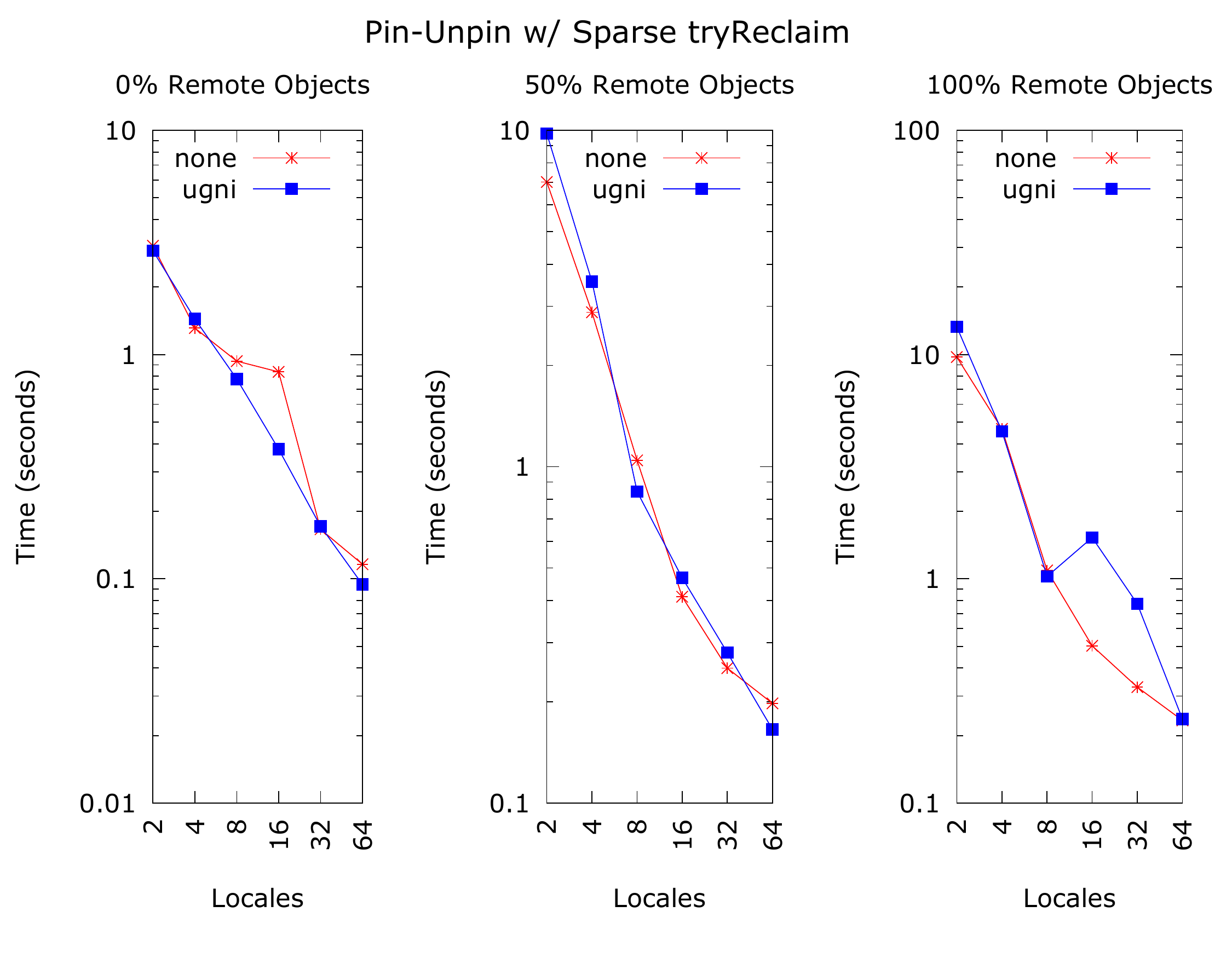}
    \caption{Deletion with \texttt{tryReclaim} called once per 1024 iterations.}
    \label{fig:sparse}
\end{figure}

\begin{figure}
    \centering
    \includegraphics[width=0.45\textwidth]{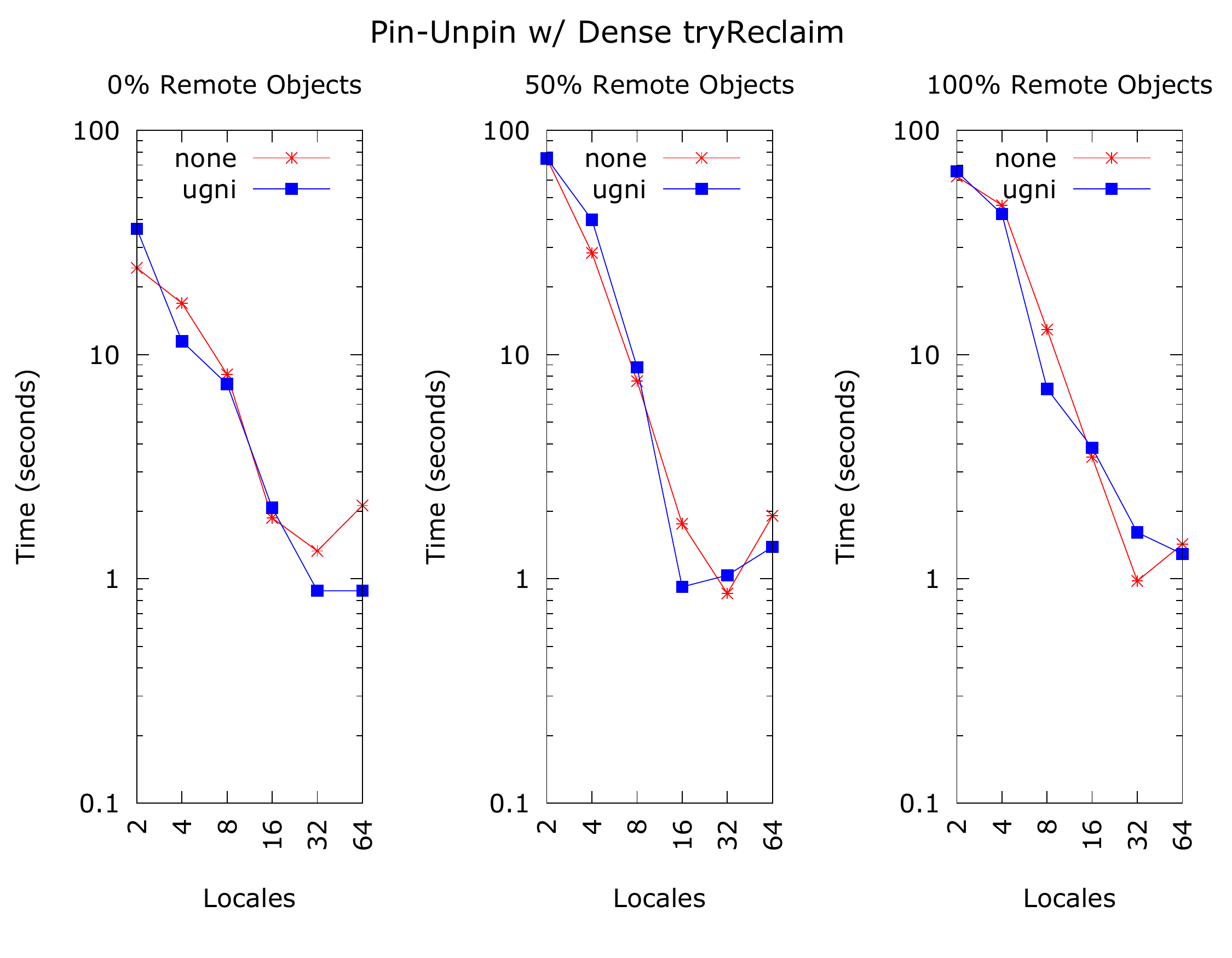}
    \caption{Deletion with \texttt{tryReclaim} called every iteration.}
    \label{fig:dense}
\end{figure}

\begin{figure}
    \centering
    \includegraphics[width=0.45\textwidth]{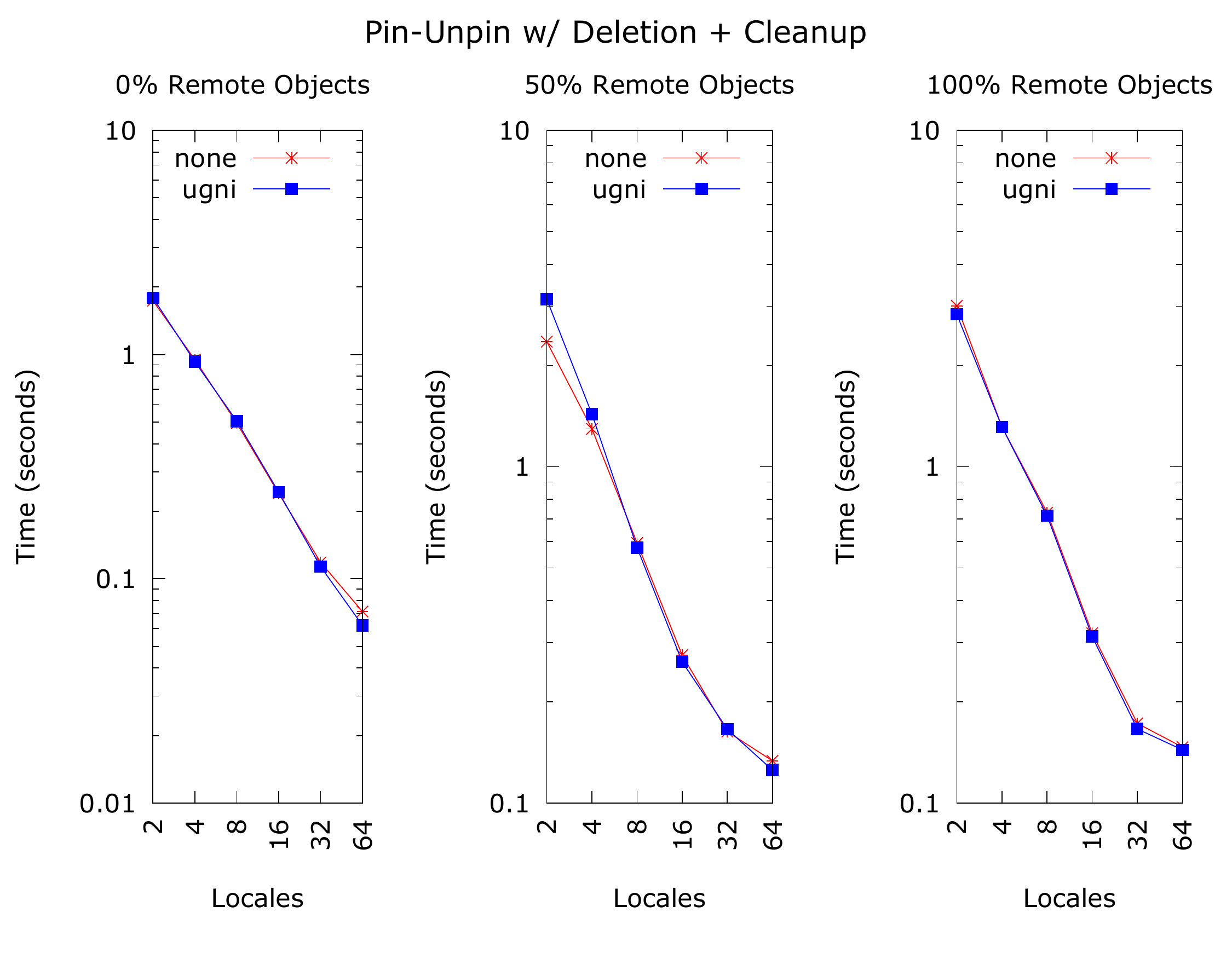}
    \caption{Deletion with reclamation only performed at end.}
    \label{fig:cleanup}
\end{figure}

\begin{figure}
    \centering
    \includegraphics[width=0.45\textwidth]{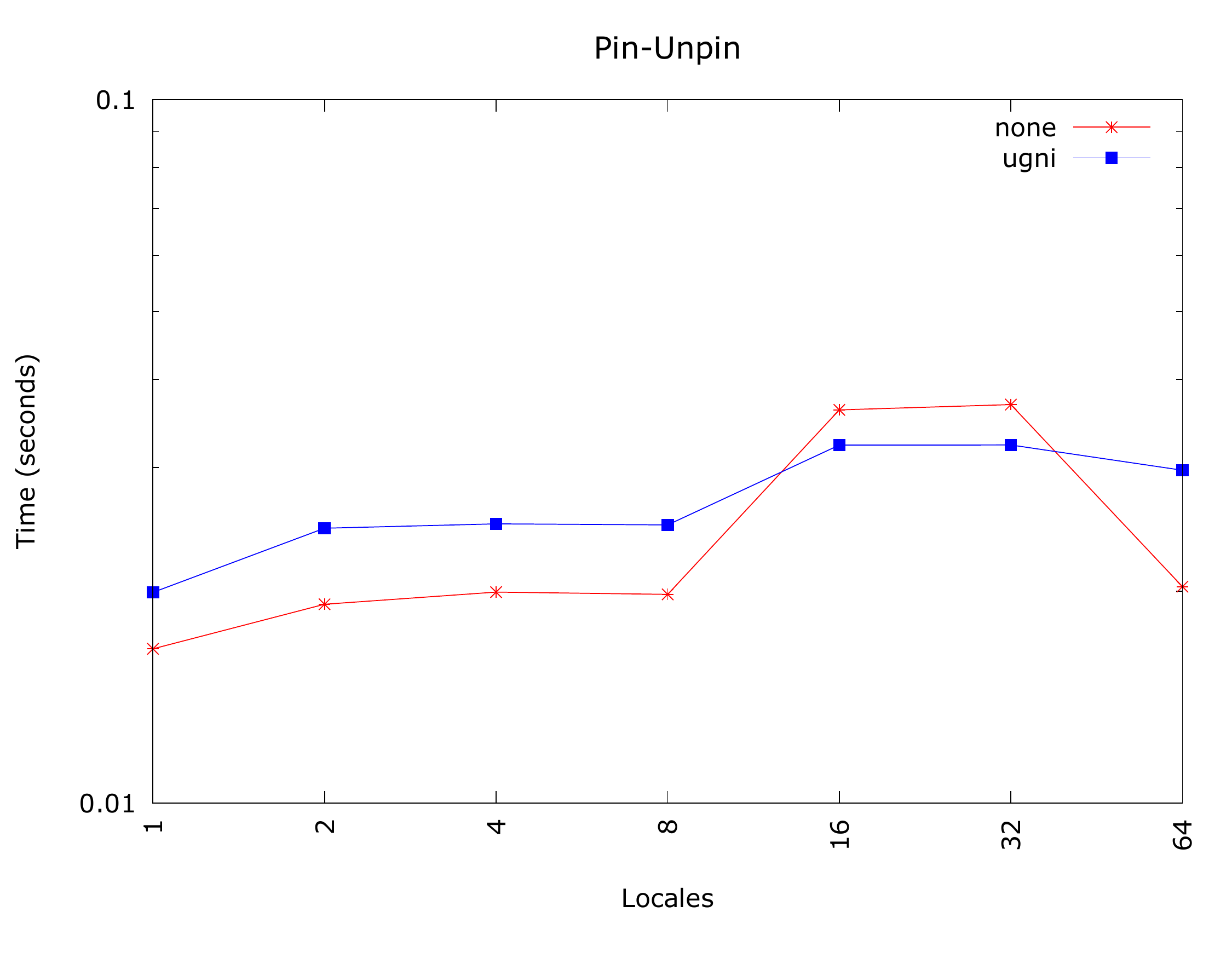}
    \caption{Read-only workload without deletion.}
    \label{fig:raw}
\end{figure}

All experiments were conducted on a 64-node Cray XC-50 with 44 core Broadwell CPUs per node, compiled using Chapel 1.20 with the `\-\-fast' flag to enable all compiler and backend optimizations. The experiments were conducted both in the presence and absence of \texttt{CHPL\_NETWORK\_ATOMICS},  which is RDMA atomic operations. These RDMA atomics are not coherent, and so all atomic operations, including those that are performed locally on the same system, must go through the Gemini or Aries NIC. This overhead of using network atomics for local operation has been measured to be as much as an order of magnitude by the authors. While in the development of both the \texttt{EpochManager} and \texttt{AtomicObject}, care has been taken to eliminate the usage of RDMA atomics when they are unnecessary by `opting out', it is still useful to compare performance with and without the support for RDMA atomics. Since RDMA atomics are only available on Gemini and Aries, the performance observed with RDMA atomics disabled would be relatively close to the performance seen when using InfiniBand, as Chapel does not utilize InfiniBand RDMA atomics even when present on the system.

The performance of both \texttt{AtomicObject} and \texttt{EpochManager} is measured to show the raw overhead of both constructs with the goal of proving them both scalable. Such microbenchmarks are important in that it is exceedingly difficult to build scalable non-blocking algorithms without scalable building blocks, beginning with \textit{AtomicObject}, which is a fundamental building block, including for \texttt{EpochManager}. The \texttt{AtomicObject} is compared to one of the only types that atomics are natively supported in Chapel, the \texttt{atomic int}. The \texttt{atomic int} is also a sibling of the \texttt{atomic uint}, which the \texttt{AtomicObject} is built on top of. Microbenchmarks involving the \texttt{AtomicObject} test the overhead injected by the abstraction. Microbenchmarks involving the \texttt{EpochManager} focus on different use-cases and workloads.

In this section, we explore two performance criteria. First, we evaluate the performance of Atomic Objects against Chapel's atomic variables. We use Chapel's \texttt{atomic int}. The experiments focus on the common set of operations available between Chapel's atomic variables and Atomic Objects: read, write, compare and swap, and exchange. Second, we evaluate the scalability of \texttt{EpochManager}, testing raw acquire/release, memory reclamation with remote objects, and manual garbage collection every fixed number of iterations.

\subsection{Atomic Objects Performance Evaluation}

We compare Atomic Objects performance with and without ABA protection against Chapel's \texttt{atomic int} in shared memory and distributed memory, as shown in Figure~\ref{fig:atomics}. The experiment evaluates strong scaling, with each task performing the same number of operations, comprising 25\% read, 25\% write, 25\% compare-and-swap, and 25\% exchange operations. Shared memory experiments show that all three of atomic int, AtomicObject (ABA) and AtomicObject scale linearly with an increasing numbers of tasks. AtomicObject without ABA protection performs equivalently to Chapel's atomic int, and AtomicObject (ABA) takes the highest amount of time, with a constant overhead. In distributed memory, the performance of AtomicObject without ABA protection is equivalent to Chapel's atomic int. This shows that even in distributed memory, there is no noticeable overhead, and it scales linearly with the number of locales, whether RDMA atomics are used or not. AtomicObject (ABA) scales linearly with an increasing numbers of locales. It performs equivalently to Chapel's atomic int without network atomics.

\subsection{Epoch Manager Performance Evaluation}

\begin{listing}
\begin{minted}[fontsize=\scriptsize]{chapel}
// Create manager instance
var manager = new EpochManager();
var objsDom = {0..#numObjects} dmapped Cyclic(startIdx=0);
var objs : [objsDom] unmanaged C();
// Randomize locale that each object is allocated on
randomizeObjs(objs);
forall obj in objs with (
  var tok = manager.register(), 
  var M : int
) {
  tok.pin();
  // If we are deleting...
  tok.deferDelete(obj);
  tok.unpin();
  M += 1;
  // If we are tryReclaim'ing...
  if M % perIteration == 0 {
    tok.tryReclaim();
  }
}
// Reclaim all objects at end
manager.clear();
\end{minted}
\caption{Microbenchmark of \texttt{EpochManager}.}
\label{lst:micro}
\end{listing}

The microbenchmarks for \texttt{EpochManager} are similar to Listing~\ref{lst:micro}.

We evaluate the scalability of \texttt{EpochManager} under various workloads, which should be representative of the different use-cases. In a read-only workload, such as for a read-often write-rarely data structure, such as when performing a lookup in a hash table or a linked list, it may be suitable to just pin at the beginning of the operation, and then unpin at the end. Demonstrated in Figure~\ref{fig:raw}, performance is essentially stable across multiple locales, demonstrating that even in distributed contexts it can scale reasonably well as all locales forward their accesses to their privatized instances despite being in a parallel and distributed \texttt{forall} loop. In Figure~\ref{fig:cleanup}, another typical workload is analyzed where no reclamation is performed until the very end, which is typical when the number of objects is bound and can fit within memory without running out of memory. The number of remote objects to be reclaimed varies by 0\%, 50\%, and 100\%, which measures the overhead of reclaiming remote objects.
The \texttt{EpochManager} scales even in the case where \texttt{tryReclaim} is invoked with increasing frequency, as demonstrated by the results displayed in Figure~\ref{fig:sparse}. When reclamation is performed, not even the locale where the global epoch is allocated is bogged down by redundant requests thanks to the first-come-first-serve election of tasks, and scales equally both with and without RDMA atomics. In the case where the user does not want to take any chances and attempts to \texttt{tryReclaim} on every iteration, there is still scalability, as shown in Figure~\ref{fig:dense}.

\section{Conclusion}


The \texttt{AtomicObject} is a solution to the problem of a lack of language support atomic operations on objects. The implementation not only provides the operations in shared-memory but distributed memory, utilizing pointer compression that enables RDMA atomic operations, which are on-par with the performance for atomic operations on integers, while also providing protection from the ABA problem with memory reclamation via usage of double-word compare-and-swap. The \texttt{EpochManager} is a non-blocking epoch-based reclamation garbage collection system that allows for concurrent-safe reclamation even in distributed-memory contexts. Both of these are essential building blocks for developing non-blocking algorithms in both shared-memory and distributed-memory. In future works, it is planned to allow more than $2^{16}$ locales while still allowing RDMA atomic operations, by introducing another level of indirection and utilizing an descriptor index into a separate table of objects in place of the pointer itself. As well, there is planned exploration of allowing atomics on owned and borrow types. Also in future works, an application of both the constructs in the porting of the Interlocked Hash Table~\cite{jenkins2017redesigning} is complete and awaiting release; their applications in the creation of other distributed algorithms are also planned.

\bibliographystyle{ieeetr}
\bibliography{references}

\end{document}